 \documentclass[twoside,reqno,11pt]{gocp}

\usepackage{graphicx}
\usepackage{epsfig}
\usepackage{amsthm}
\usepackage{amsmath}
\usepackage{latexsym}
\usepackage{amsfonts}
\usepackage{amssymb}

 \usepackage{hyperref}

 \def\theequation{\arabic{section}.\arabic{equation}}

 \title[On the origin of space  \dots]{
On the origin of space}
 \author[R. Herrmann]{Richard Herrmann $^1$}

\begin{document}
\noindent
\includegraphics[scale=0.15]{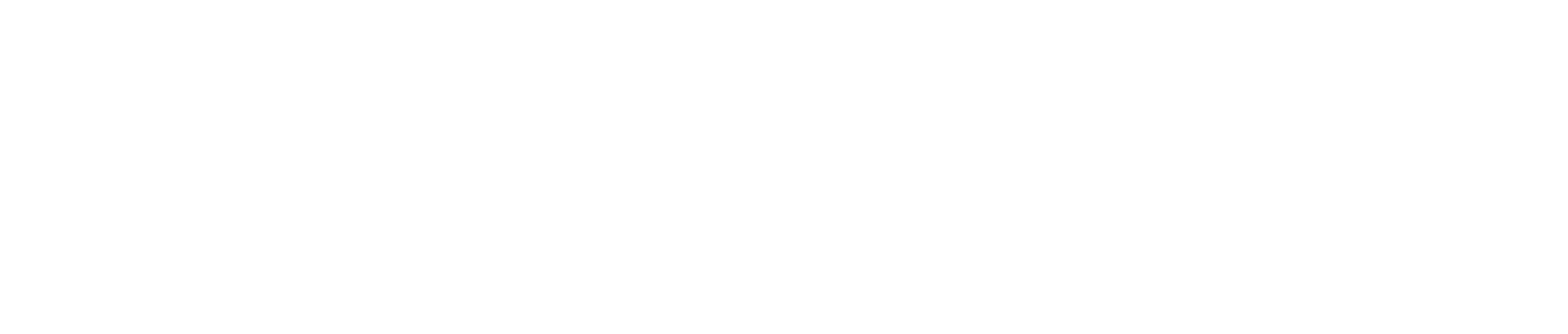}

\bigskip
\bigskip 

 \begin{abstract}
Within the framework of fractional calculus with variable order the evolution of space in the adiabatic limit is investigated. Based on the Caputo definition of a fractional derivative using the fractional quantum harmonic oscillator a model is presented, which describes space generation as a dynamic process, where the dimension $d$ of space evolves smoothly with time in the range $0 \leq d(t) \leq 3$, where the lower and upper boundaries of dimension are derived from first principles.  It is demonstrated, that a minimum threshold for the space dimension is necessary to establish an interaction with external probe particles.  A possible application in cosmology is suggested.
 \medskip

{\it Key Words}: Fractals and nonlinear dynamics, Quantum mechanics,  Schr\"odinger equation, Cosmology
 \smallskip

{\it PACS}:  05.45.Df, 03.65. -w, 02.60.Lj
 \end{abstract}

 \maketitle

\section{Introduction}
Fractional calculus introduces the concepts of nonlocality and memory in order to model complex dynamical behaviour \cite{f3, sam93a, pod99, hil00, he11}. Literally spoken, we have learned, that the present is influenced by the past in more than one way. Within the framework of fractional calculus it really makes a difference how an actual state is achieved.  

Alternatively, from a historical point of view  fractional calculus may be described as an extension of  the concept of a derivative operator from integer order $n$ to arbitrary order $\alpha$, where $\alpha$ is a real or complex value \cite{f1}:
\begin{equation}
\label{c1first}
{d^n \over dx^n} \rightarrow {d^\alpha \over dx^\alpha}
\end{equation}
In this talk we want to introduce a further facet, which emerges naturally from a new point of view based on fractional calculus with variable order. It is the fascinating idea of an evolutionary development of dynamic processes, which are described by differential equations, which themselves evolve dynamically in evolutionary scenarios; an idea, that dates back to the early 90s of the last century \cite{sam93a,ram10, odz11,odz13} and which may be  realized introducing a time and/or space dependence for $\alpha$.
\begin{equation}
\label{c26second}
\alpha \rightarrow \alpha(x, t)
\end{equation}
\index{genetic fractional differential equations}
\index{evolutionary fractional differential equations}
The governing differential equations, which until now remained form invariant within an ever changing environment, now themselves take part in  a global evolutionary process.  

With respect to  this concept, we will also take a look into the future of fractional calculus and we anticipate developments that have yet to be achieved. 
The time has come to present an unconventional approach to the still open question of all questions, on the origin and the evolutionary genesis of space.

\setcounter{equation}{0}
\section{The interplay between matter and space}

In times of Aristotle matter and space were interdependent quantities, where one could not exist without the other. This holistic view allowed an understanding of phenomena like {\it{horror vacui}} and gave a direct interpretation of e.g. von Guericke's experiments \cite{gue72} in the second half of the 17nth century. 

As a fundamental assumption in Newtonian physics the concept of matter was established as an independent quality in an absolute space and time, which left themselves reduced to passive containers of a dynamically developing material world.
\index{horror vacui}

This tendency to interpret space as a passive container for  dynamically interacting  particles becomes most obvious in quantum theory, where the vacuum state, which initially was characterized by the absence of any particles, since the days of Dirac,  serves more and more as a waste deposit for  dispensable virtual matter of all kind.

In Einstein's theory of general relativity the hitherto disjunct concepts of space, time  and matter are once again connected via the field equations
\begin{equation}
R_{\mu \nu} -  \frac{1}{2} g_{\mu \nu } R = \kappa T_{\mu \nu} 
\end{equation}
which relate a property of space-time  called curvature with a property of matter collected in the energy-momentum tensor. In the case of matter absent, these equations reduce to the vacuum field equations
\begin{equation}
R_{\mu \nu} -  \frac{1}{2} g_{\mu \nu } R = 0 
\end{equation}
or equivalently $R_{\mu \nu}=0$,
\index{Einstein's filed equations}
which generates a flat space.

Hence in Einstein's field theory  besides curvature there is no other property of space affected. Especially the number of space and time dimensions is invariant (3+1) under any Lorentz-transformation, which results in a very static picture of space at the end.

The first serious attempt to overcome the paradigm of  a fixed number of space dimensions  was realized by Kaluza \cite{kal19} and Klein \cite{kle26} introducing a 5-dimensional theory of relativity, which allowed for a simultaneous treatment of gravity and electro-magnetism. 

The Kaluza-Klein theory may be considered as one of the ancestors of string theory, which gained increasing interest, after it was shown, that a higher dimensional realization introducing extra dimensions avoids divergences in a  perturbation expansion. Of course, to be comparable with experimental observations, a mechanism to reduce a multi-dimensional theory to a 3-dimensional in space had to be introduced as compactification.
\index{compactification}
\index{Kaluza-Klein theory}

Despite a few   occasional attempts \cite{tan63, he90, ell92, cal11}, up to now there is no generally accepted theory, which explains the observed number of space dimensions and Ehrenfest's \cite{ehr17}  question, why space has 3 dimensions, is still unanswered. Furthermore, there is no dynamic theory, which rationalizes the genesis of 3-dimensional space. 

Instead e.g. the classical big bang model starts with an object with very small size but right from the beginning the number of space dimensions is fixed to a value, which remains unchanged till today.  

At this point fractional calculus comes into focus as a possible candidate to overcome some of the restrictions of hitherto discussed approaches.

We will collect arguments, that space and especially the space dimension $d$ is a dynamic quality. Considering space as a quantum phenomenon, it evolves smoothly from a point-like vacuum fluctuation with $d=0$ up to it's final form $d=3$, where d is a real number. Therefore interpreting $d \in \mathbb{R}$ instead of $d \in \mathbb{N}$ seems a necessary prerequisite and is a natural legitimation for the use of a fractional concept.  This is our first argument for a fractional approach. 

Furthermore space dimension is considered a dynamic quality $d=d(t)$, which clearly emphasizes the evolutionary aspect of space genesis. This is the second argument for a fractional approach and in the following we will model this behaviour by investigating the properties of a fractional wave equation with time dependent $\alpha=\alpha(t)$. 

Besides the known mechanisms for an explanation of an inflating universe, we propose in the following  an evolutionary approach, where space is permanently generated starting from a dimensionless infinitely small seed up to its final $d=3$ form which may be compared to  gas bubbles in boiling water.

\setcounter{equation}{0}
\section{Fractional calculus with time dependent $\alpha$ in the adiabatic limit}

\begin{figure}
\begin{center}
\includegraphics[width=125mm]{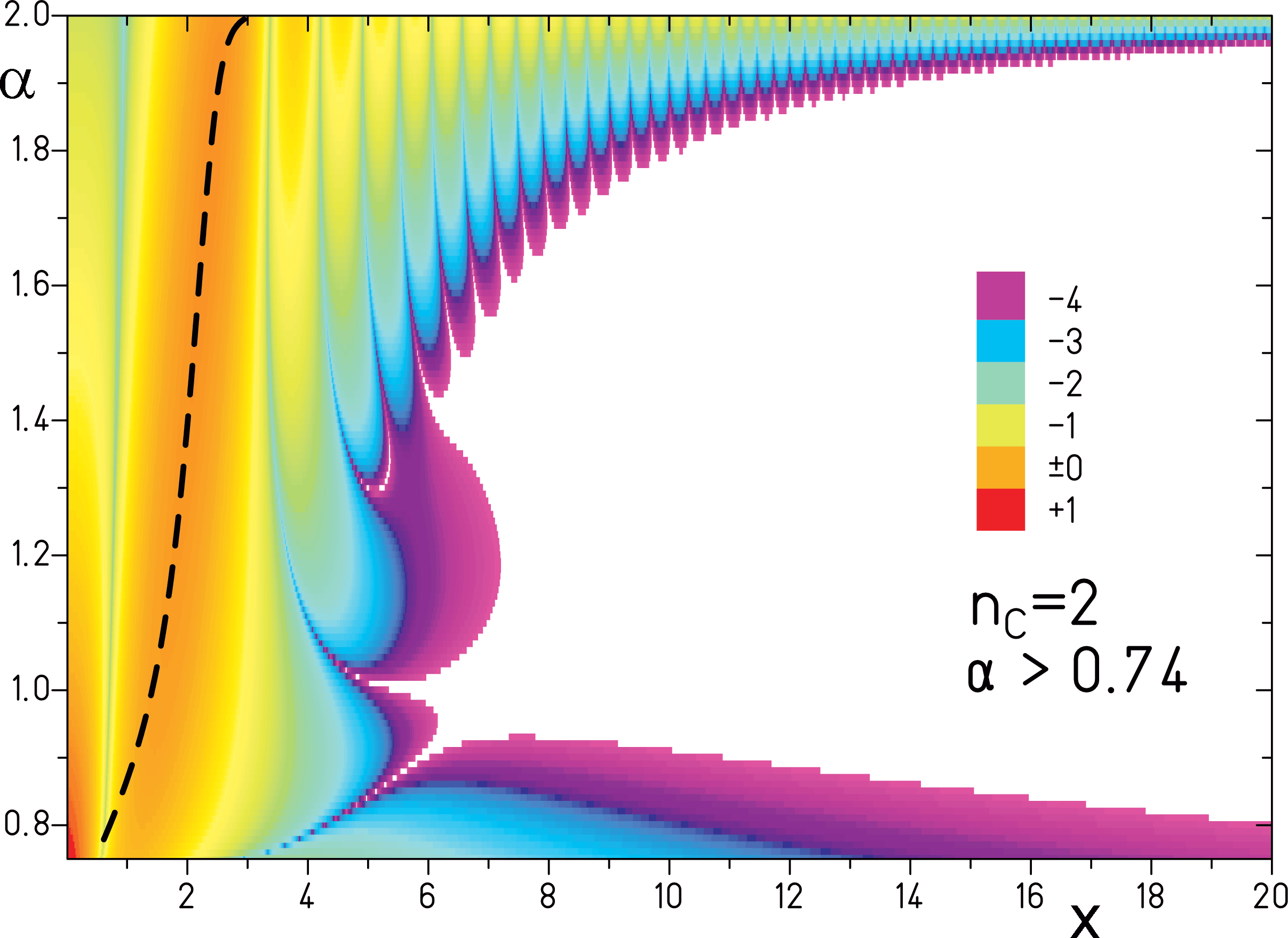}\\
\caption{\label{both26}
Logarithmic probability density  $|\Psi_{n=2}^+(x(t),\alpha(t))|^2$ for the solution of the fractional quantum harmonic oscillator based on the Caputo definition of the fractional derivative in the full range of allowed $\alpha$ values for real energy eigenvalues. The dashed line indicates the expectation value for the size operator  $\bar{x}=H(x) x$ ($H(x)$ is the Heaviside step function), its expectation value
$
\langle \bar{x} \rangle = 
\int_{0}^\infty dx \Psi_n^\pm(x,\alpha) \, x \, \Psi_n^\pm(x,\alpha)^* 
$
yields the position information on the positive semi-axis.   
} 
\end{center}
\end{figure}

Since we consider the genesis of space as a quantum phenomenon, we describe this process with the simple model of the fractional quantum harmonic oscillator \cite{he12a}:
\begin{equation}
\label{f26ho1}
H^{\alpha(t)} \Psi(x, t) =(  {1 \over 2 }\bigl( - (_{\textrm{\tiny{C}}}\hat{D}^{\alpha(t)}_{x})^2+ (|x|^{\alpha(t)})^2 \bigr) \Psi(x, t) =  - i \hbar \partial_{t}\Psi(x, t)
\end{equation}
using the Caputo fractional derivative $_{\textrm{\tiny{C}}}\hat{D}^{\alpha}_{x}$, which is related to the Riemann fractional derivative $_{\textrm{\tiny{R}}}\hat{D}^{\alpha}_{x}$ via, \cite{he11,tar07}:
\begin{eqnarray}
\label{c26cc}
{_\textrm{\tiny{C}}}\hat{D}^{\alpha}_{x} &=& {_\textrm{\tiny{R}}}\hat{D}^{\alpha-1}_{x} \partial_x \\
    &=& {1
\over
\Gamma(2-\alpha)}
|x|^{1-\alpha}
:\!{_1}F_1(1-\alpha;2-\alpha;-x \partial_x)\!: \partial_x 
\end{eqnarray}
where the hypergeometric function ${_1}F_1(a;b;z)$ is interpreted   as a series expansion in terms of integer derivatives, where $:(x \partial_x)^n: = x^n \partial_x^n$ is the normal ordered product and 
with time dependent $\alpha(t)$, where the set $\{x,t\}$  marks two different space- and time-like variables.

In order to simplify the procedure, we assume complete adiabaticity, which means, that the system has always time enough to occupy the  equilibrium state. In that case, we may treat the time $t$ as a parameter and the problem is reduced to the solution of  a one dimensional stationary Schr\"odinger equation in $x$, where $t$ parametrizes the solutions.

\begin{figure}[t]
\begin{center}
\includegraphics[width=85mm]{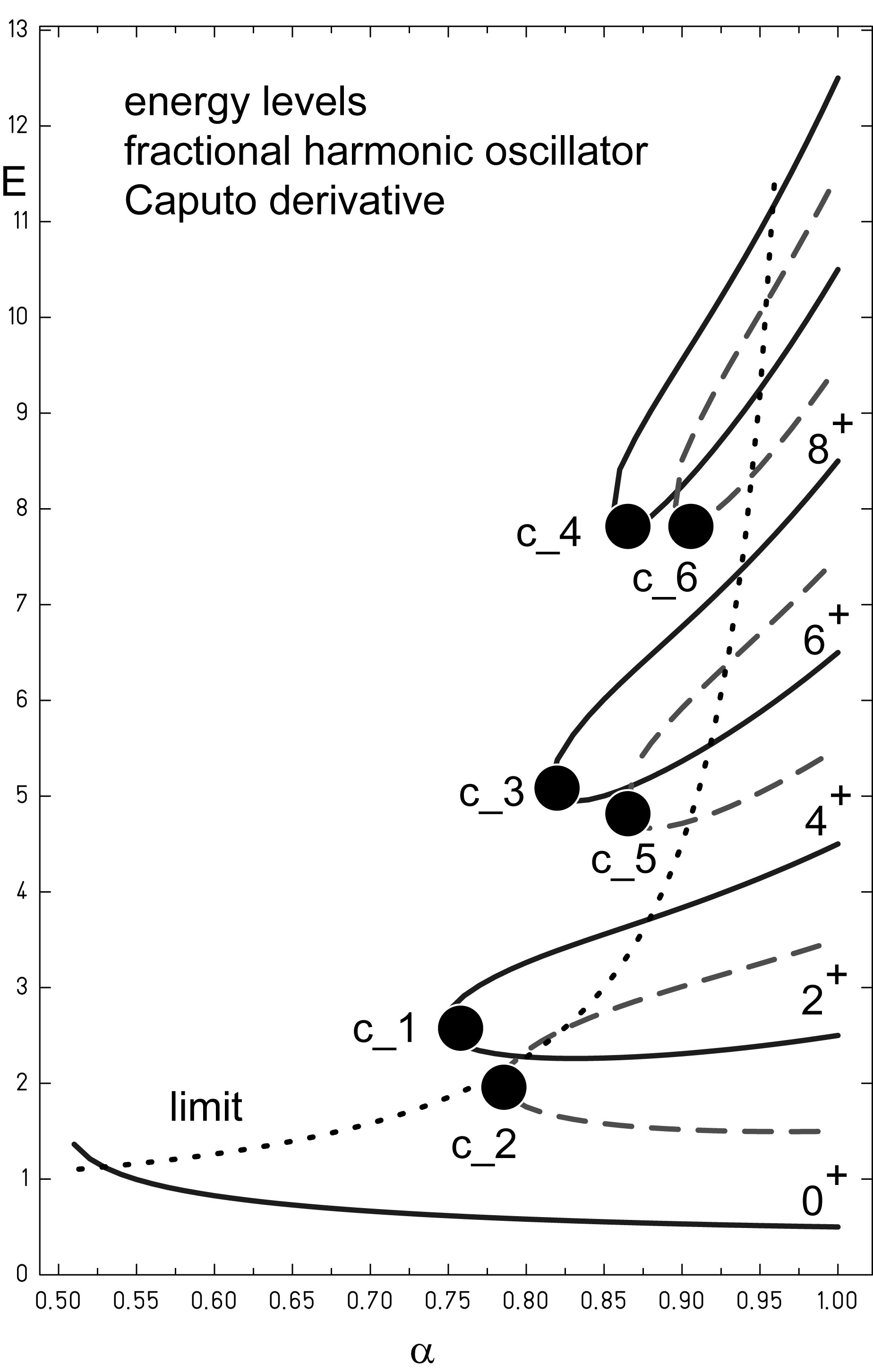}\\
\caption{\label{energy levels HOC}
The energy spectrum for the fractional quantum harmonic oscillator using the Caputo derivative. Thick lines indicate positive parity, dotted lines indicate negative parity. Bullets depict the bifurcation points ($c_i$) determined numerically (see table  \ref{tabhoClimits}). Pointed line according to (\ref{xaaq}) shows a first crude guess for bifurcation points from the requirement, that the occupation probability for a given state should be a positive number.
} 
\end{center}
\end{figure}

An approximate solution for the stationary energy levels, which is valid in the neighbourhood of $\alpha \approx 1$  has been derived by Laskin  \cite{laskin} within the framework of WKB-approxi\-ma\-tion, 
\index{WKB-approximation}
which is independent from a specific choice of a fractional derivative type:
\begin{equation}
\label{ehowkb11}
E_{\textrm{\tiny{WKB}}}(n,\alpha)  =  
\bigl(
n + {1 \over 2}
\bigr)^\alpha
\pi^{\alpha/2}\left({\alpha \Gamma(\frac{1+\alpha}{2 \alpha}) \over \Gamma(\frac{1}{2 \alpha})}\right)^\alpha  
\, n=0,1,2,...
\end{equation}
We have already emphasized \cite{he12a}, that these levels allow for a smooth transition from vibrational to rotational types of spectra, depending on the value of the fractional derivative coefficient $\alpha$. 
\begin{eqnarray}
\label{ewkbapprox11}
E_{\textrm{\tiny{WKB}}}(n,\alpha \approx 1) & \sim&  
n + {1 \over 2}  \quad   \quad    \quad  n=0,1,2,...\\
E_{\textrm{\tiny{WKB}}}(n,\alpha \approx 2) & \sim&  
 (n+{1 \over 2})^2 =  n (n+1) + 1/4
\end{eqnarray}
Indeed the solutions of the fractional quantum harmonic oscillator cover vibrational as well as rotational degrees of freedom from a generalized view as fractional vibrations \cite{he103}.

Exact eigenfunctions and eigenvalues have been obtained numerically recently \cite{he13a}. As an example, in figure \ref{both26} we have plotted a typical eigenfunction ($\Psi_{n=2}^+(x,\alpha)$). The eigenfunctions, normalizable only for $\alpha<2$,
determine a Hilbert space and are used to calculate a measure  of the  time dependent size $s/2 = \langle \bar{x} \rangle $ given in terms of  the modified position expectation value
\begin{equation}
\label{sizexHO}
\langle \bar{x}(t) \rangle = \langle H(x(t)) x(t) \rangle=
\int_{0}^\infty dx \Psi_n^\pm(x,\alpha(t)) \, x \, \Psi_n^\pm(x,\alpha(t))^* 
\end{equation}
where $H(x)$ is the Heaviside step function. 

\begin{table}
\begin{center}
\caption{Bifurcation points ($c_i$) corresponding to the lowest excited states for the energy levels of the fractional quantum harmonic oscillator using the Caputo derivative. Points are sorted for increasing $\alpha$ and determined as the lowest real energy $E_c$ for a given $\alpha=\alpha_c$.}
{\begin{tabular}{r|rrr}
\hline\noalign{\smallskip}
$c_i$& levels  & $\alpha_c$ & $E_c$\\
\noalign{\smallskip}\hline\noalign{\smallskip}
1&$2^ + - 4^+$      & 0.7463952 &   2.6065\\
2&$1^-   - 3^-$       & 0.7847781  &  2.0067\\
3&$6^ + - 8^+$      & 0.8156782 &  5.1369\\
4&$10^ + - 12^+$   & 0.8501847 &  7.8612\\
5&$5^-   - 7^-$       & 0.8603896 &   4.9009\\
6&$9^-   - 11^-$      & 0.8919590 &  7.9842\\
\noalign{\smallskip}\hline\noalign{\smallskip}
\end{tabular}}
\label{tabhoClimits}
\end{center}
\end{table} 

The time development of a space element with size $s$ starts for $\alpha=0.5$ with $s(\alpha=0.5) = 0$ for the 
ground state $\Psi_0^+$ and above $\alpha=1$ it increases slowly, but becomes divergent for $\alpha>2$. Therefore when $\alpha$ becomes time dependent,  the expectation value $\langle \bar{x}(t) \rangle $ may then be considered as a measure of the size development of a space element with time and may be compared to a bursting soap bubble. 

In figure \ref{energy levels HOC} we present the exact energy spectrum for the fractional quantum harmonic oscillator using the Caputo derivative definition for the range $0.5 \le \alpha \leq 1$. 

For $\alpha<1$ we obtain only a limited number of eigenvalues. In the limiting case $\alpha = 0.5$ we are left with only a single energy level. 

Based on the stationary fractional quantum harmonic oscillator the infrared spectrum of HCl has been reproduced successfully \cite{he12a}. It has also been demonstrated, that the transition from vibrational ($\alpha \approx 1$) to rotational type ($\alpha \approx 2$) of spectra may be interpreted within the context of the fractional quantum harmonic oscillator as a transition from one- $d\approx1$  to three-dimensional $d\approx 3$ space, as long as the corresponding multiplicities are correctly implemented \cite{he12a}. 

Therefore in the following  we will consider the consequences of a time dependence  of $\alpha=\alpha(t)$, which implies a time development of the space dimension $d=d(t)$:  
\setcounter{equation}{0}
\section{The model and possible consequences for an application in cosmology}
We know, that the infrared spectra of diatomic molecules may be reproduced based on the level spectrum of the fractional quantum harmonic oscillator by defining the relative                                        intensities as a Boltzmann distribution weighted with the degeneracy of the n-th energy level \cite{he12a}, where $\beta = 1/(k_B T)$:
\begin{equation}
I_n = (2(\alpha -1)n+1)   e^{-\beta E_n}
\end{equation}
Which clearly demonstrates the unifying aspects of the fractional ansatz, since it combines the standard vibrational as well as the standard rotational elements as special cases
\begin{eqnarray}
I_n^{\textrm{vib}} &=&  e^{-\beta (n+1/2)}, \,\qquad\qquad\quad \alpha=1 \\
I_n^{\textrm{rot}} &=&  (2 n+1) e^{-\beta n (n+1)}, \qquad \alpha=2 
 \end{eqnarray}
if we associate $n$ with the vibrational quantum number for the standard one-dimensional quantum harmonic oscillator and with the total angular momentum quantum number $L$ with degeneracy $2 L +1$ for rotations in $\mathbb{R}^3$. 

For a practical application,  the  spectra of diatomic molecules are reproduced with high accuracy setting $\alpha \approx 2$ for rotational spectra and $\alpha\approx 1$ for vibrational spectra and therefore prove the validity of the fractional approach in the close vicinity of $\alpha = 1,2$ only. 

Now we want to investigate this approach for the full range of allowed $\alpha$-values, especially $\alpha \ll 1$ and parametrize with time $t$.   

In order to calculate the time dependent spectral response of a space element, we therefore postulate that the dimensionality of a given space element is directly related to the multiplicity of the occupation probability $p^{\alpha(t)}_n $  for a given state $\psi_n(x,t)$.
\begin{equation}
p^{\alpha(t)}_n = m(d(t),n) \frac{1}{Z^{\alpha(t)}} e^{-\beta E_n^\alpha(t)}
\end{equation}
as a product  of a thermalized level distribution with a dimension dependent multiplicity factor  $m(d(t),n)$, which determines the dimension dependent degeneracy of the n-th eigenvalue  and 
with the extended canonical ensemble $Z^{\alpha(t)}$ 
\begin{equation}
Z^{\alpha(t)} = m(d(t),n) \textrm{Tr}(e^{-\beta H^\alpha(t)})
\end{equation}
where $\beta = 1/(k_B T)$ and the multiplicity $m$ defined as
\begin{equation}
\label{mm}
m = (d(t)-1) n + 1 
\end{equation}
where the connection between dimension $d$ and the fractional parameter $\alpha$ is given as:
\begin{equation}
\label{spdima}
d(t) = 2 \alpha(t)-1, \quad \quad {{1} \over{2}}<\alpha \leq 2
\end{equation}
Indeed, with these settings,  for the special  case $d \in \mathbb{N}$ we obtain the multiplicities
\begin{eqnarray}
m(d=1) &=& 1\\
m(d=2) &=& n+1\\
m(d=3) &=& 2 n +1
\end{eqnarray}
which is conformal with classical quantum mechanics, if we agree, that the fractional quantum harmonic oscillator  bridges the two different Lie algebras $U(1) \rightarrow O(3)$ and $n$ counts the multiplets of the corresponding Casimir-operators $\hat{N}$ and $\hat{L}^2$   as outlined in e.g. \cite{he103, he12a}. By the way, the explicit properties of the  group elements and Casimir-operator for $d=2$  have not been investigated yet.

Consequently, we obtain a first natural explanation for the upper limit $d=3$ of allowed space dimensions from the requirement of normalizability of the solutions of the fractional quantum harmonic oscillator, which requires $\alpha<2$. This is a fascinating answer to Ehrenfest's question \cite{ehr17}.   
 
Equation (\ref{mm}) also leads to a natural explanation for the limited number of levels for small $\alpha$:
From the requirement that  the multiplicity  of a given state $n$ should be  a positive real number, $m>0$, which means, within the framework of a possible particle-hole, or better space-anti-space formalism (since we discuss a property of space, not matter) 
\index{particle-hole formalism}
\index{space-anti-space formalism}
 we restrict to a description of space  only,  using (\ref{mm}) a condition follows
for the  finite set of allowed n values $n \in \{0,...,n_{\textrm{\tiny{max}}}\}$: 
\begin{equation}
\label{xaaq}
n{\textrm{\tiny{max}}} < {{1} \over {2(1-\alpha(t))}},  \quad \quad \alpha < 1
\end{equation}
which in the limiting case  $\alpha \rightarrow {{1} \over{2}}$ reduces to only one allowed level with $n=0$. 

Consequently we obtain a natural explanation for the lower limit of allowed space dimension, since below $\alpha=0.5$ there are no solutions for the fractional quantum harmonic oscillator, which according to (\ref{spdima}) leads to the lower limit of $d=0$.

In figure  \ref{energy levels HOC} this limiting function (\ref{xaaq}) is plotted with pointed lines and it shows a qualitative agreement with the exact data.
\begin{figure}
\begin{center}
\includegraphics[width=120mm]{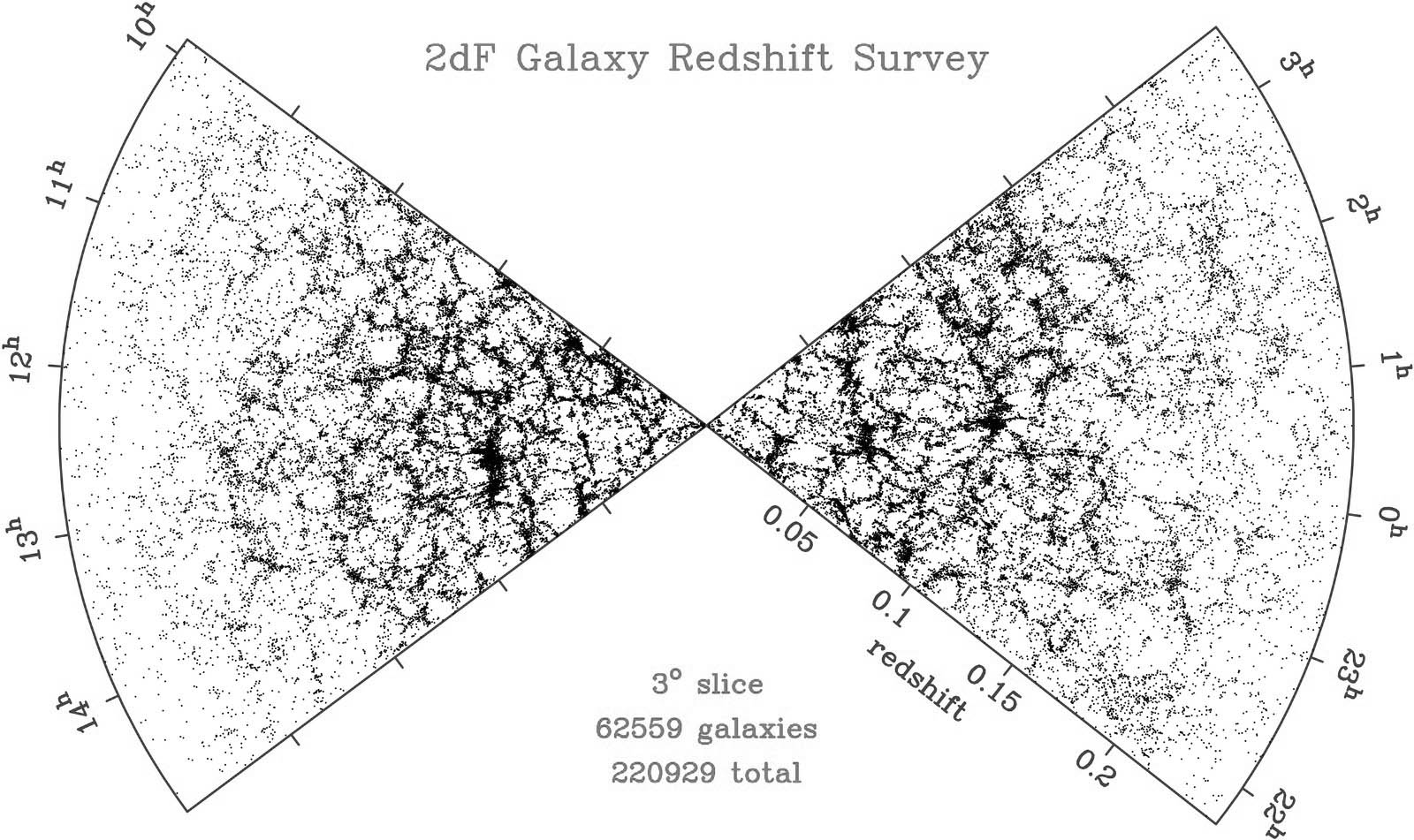}\\
\caption{Results of the 2dF redshift galaxy survey. Here one of the redshift slices for different spectral types is shown. It is a graphical presentation of our interpretation of space genesis and inflation of the universe. Space dominated regions are considered as matter free and shown as white areas, while matter dominated areas may be considered as boundaries and are presented in dark colour. In a classical approach,  space dominated regions 
may be considered as dominated by dark energy $E_d$ while matter dominated regions may be interpreted as a surface of a space generating region, emulating a compressive contribution which may be interpreted classically as the influence of a dark matter $M_d$. 
 Credits go to the "the 2dF Galaxy Redshift Survey team".
 With kind permission, see Colless et al. [4]
source from:  
http://www2.aao.gov.au/2dFGRS 
\label{galaxy survey}} 
\end{center}
\end{figure}

From the time dependence of $\alpha = \alpha(t)$ and the assumed correspondence of $\alpha$ with the space dimension $d$ according to (\ref{spdima}) we now propose an evolution of the space dimension as a function of time which is mediated by e.g. 
\begin{equation}
\label{xaaq2}
\alpha(t) = \frac{1}{2}+\frac{3}{2}  \arctan(t)  \quad 0 \leq t \leq \infty
\end{equation}
The expansion of the universe may be  explained within the framework of this model: In an initially mainly homogeneous environment space elements with an initial space dimension $d=0$ are seeded,  which evolve with time to $d=3$ standard space dimension, generating a foamy structure of the universe (see (\ref{sizexHO}) for a size estimate as a function of $\alpha(t)$).

In figure \ref{galaxy survey} the distribution of space and matter for a small slice of space shows exactly this foamy structure.
Space dominated regions are considered as matter free, shown as white areas, while matter dominated areas may be considered as boundaries and are presented in dark colour. The white regions are therefore permanent sources of space generation and cause a major contribution to the expansion of the universe.

The space dimension may be probed experimentally by a measurement of an intensity distribution of light as a function of an incident light ray:

From figure \ref{energy levels HOC} we may deduce the bifurcation points for a given $\alpha_c$ giving the onset of a real eigenvalue for increasing $\alpha$. In table \ref{tabhoClimits} these values are listed for increasing $\alpha$. 

Through the genesis of space starting with $\alpha = 0.50, d=0$ we therefore have a threshold space dimension of about $\alpha_c = 0.7463952$, $d_c=0.49279$, which is the minimum space dimension, allowing for excitation ($0^+ \rightarrow 2^+)$ and radiation ($2^+ \rightarrow 0^+)$ processes. Below this threshold this space element is dark.

The threshold dimension $d_c$ may be considered as 
\begin{equation}
d_c = 1/2 - \textrm{perturbation}
\end{equation}
Within that context, we want to make the following remark: a fractional derivative may be written as a Taylor-series of standard integer derivatives.

For the Caputo fractional derivative we have already derived (\ref{c26cc}),  therefore it will be no surprise, that the above numerically calculated threshold dimension $d_c$ introducing Sommerfeld's fine structure constant $\alpha_S\approx 1/137.036$ may be compared with the following quantity $D_c$:
\begin{eqnarray}  
D_c &=& \frac{1}{2} -\textrm{perturbation} \\
 &=& \frac{1}{2} -\alpha_S +\alpha_S^2  -\alpha_S^3 ... \\
      &=& \sum_{n=0}^\infty (-\alpha_S)^n  - \frac{1}{2} \\
      &=& \frac{1}{1 + \alpha_S} -  \frac{1}{2}\\
     &\approx& 0.492756
\end{eqnarray}  
\noindent
which agrees with $d_c$ up to $10^{-5}$. 
\noindent
This may be interpreted as an indication, 
\begin{itemize}
\item 
that the fractional quantum harmonic oscillator with the Caputo fractional derivative  {\it{per se}} describes an electro-magnetic charge type or alternatively may be used to give a very good estimate for the fine structure constant $\alpha_S$.

\noindent
As a consequence, we are directly led to the conclusion, that  electro-magnetic interaction  
is rather a property of space than of matter. Furthermore, e.g. the constancy of the  speed of light in the vacuum may be understood immediately, assuming it a property of space, instead of matter.

\noindent 
Even more generally we may speculate
\item
that the solution of a  fractional differential equation compared to the standard path integral approach used in QED and QCD respectively, already includes  higher order perturbations, which could be considered as a future research area of major importance.
\end{itemize}
At this point of discussion, we may introduce two possible candidates of common interest. 
In current cosmology there exist two miraculous contributions: dark energy and dark matter.
\index{dark energy}
\index{dark matter}

Dark energy was  first introduced as a necessary attribution for the  explanation for the accelerated expansion of the universe, deduced from the observed redshift of distant galaxies \cite{rie98}.

Dark matter has been considered to explain e.g.  the observations of Rubin et. al. \cite{zwi33, rub76a,rub76b} about the rotation velocity of stars in galaxies.

It is quite interesting, that both contributions carry the property darkness, which is just a phrase for the non-observability of these entities using standard approaches, which means e.g. non-existing emission or reflection of light.

In the above presented fractional model darkness in the early stage of space creation  is a natural outcome of a threshold space dimension, above which the minimum requirement for visibility, a single additional real eigenvalue, is fulfilled. It seems therefore tempting to associate regions of space genesis with high concentrations of dark energy $E_d$, while matter dominated regions may be considered as a manifestation of  a boundary effect with a compressive contribution which is interpreted classically as the influence of so called dark matter $M_d$.

The fractional ansatz therefore connects the concepts of dark energy and dark matter as a unified concept of space generation. In a foamy large scale structure of the universe, the empty areas are considered as sources of  expanding space and may be responsible for a concentration of matter at the boundaries of those space bubbles, which simulates the influence of an imaginary matter contribution.

The proposed fractional model also predicts a time development of the ratio of dark energy $E_d$ and dark matter $M_d$ of increasing type
\begin{equation}
\label{spdimainf}
 \lim_{t \rightarrow \infty} \frac{E_d}{M_d} \rightarrow  \infty 
\end{equation} 
since the time development of the dark energy $E_d$ within the framework of our fractional approach depends on a volume  $E_d \sim R(t)^{3 }$ and of dark matter $M_d$ depends on a surface  $M_d(t) \sim R(t)^{2}$ for large $t$, where $R(t)$ is a time dependent measure of the size of a given space generating region and consequently corresponds to the total size of the universe for a given time $t$, which is in accordance with current cosmological models.

\section{Conclusion}
In this paper the creation of space as  an evolutionary process has been  described within the framework of fractional calculus with time dependent varying order. 

A model was presented, which describes space generation as a dynamic process, where the dimension $d$ of space evolves smoothly with time within the range $0 \leq d(t) \leq 3$.

The proposed model gives a direct explanation for the dimension bounds: the lower bound of dimension results from the requirement of positive ($>0$) intensity value for an occupied energy level, while  the upper bound follows from the requirement of normalizability of the calculated eigenfunctions.

Furthermore a minimum threshold for the space dimension was deduced $d_c \approx 1/2$, which is  necessary to establish an interaction with external probe particles (e.g. light). Below this threshold, no absorption or emission processes are possible and as a consequence, this object cannot be detected with  spectral methods.

 This property is of fundamental importance for a successful application of the proposed model in cosmology: the fractional approach combines the concepts of dark energy and dark matter within the unified concept of dynamic space generation.   

Though the hitherto presented fractional approach is still a raw concept,  we wanted to demonstrate the potential of an extended fractional calculus, based on a varying fractional parameter $\alpha(t)$,  when applied to questions in cosmology. Many efforts have still to be made to obtain a reliable, consistent model of the universe.
But we may already conclude, that genetic models, based on evolutionary fractional differential equations, will play an increasingly important role in cosmology.

\section*{Acknowledgment}
We thank A. Friedrich  for useful discussions.

\smallskip
\sl
\noindent
$^1$ gigaHedron\\
Berliner Ring 80,                                                                                                                                                           
D-63303 Dreieich\\
 Germany  \\
e-mail: herrmann@gigahedron.com\\
\end{document}